\documentclass[twocolumn]{autart}

\usepackage{epsfig}

\usepackage{amssymb}

\usepackage{graphicx}

\usepackage{amsmath}

\usepackage{amssymb}

\usepackage{dcolumn}

\usepackage{bm}

\begin{document}

\begin{frontmatter}

\title{Impact Parameter Dependence of Inelasticity in  $pp$ / $p\overline{p}$ Collisions}

\author[uen]{P. C. Beggio}\ead{beggio@uenf.br}

\address[uen]{Laborat\'orio de Ci\^encias Matem\'aticas, Universidade
  Estadual do Norte Fluminense Darcy Ribeiro - UENF,\\
  Campos dos Goytacazes - RJ, Brazil.}


\begin{abstract}
We study the impact parameter dependence of inelasticity in the
framework of an updated geometrical model for multiplicity
distribution. A formula in which the inelasticity is related to the
eikonal is obtained. This framework permits a calculation of the
multiplicity distributions as well as the inelasticity once the
eikonal function is given. Adopting a QCD inspired parametrization
for the eikonal, in which the gluon-gluon contribution dominates at
high energy and determines the asymptotic behavior of the cross
sections, we find that the inelasticity decreases as collision
energy is increased. Our results predict the KNO scaling violation
observed at LHC energies by CMS Collaboration.
\end{abstract}

\end{frontmatter}

\section{Introduction}

Inelasticity is defined as the fraction of the available energy
relesead for multiple particle production in inelastic hadronic
interactions. The remaining part of the incident energy is carried
away by the participant's remnants, so called leading particles. The
energy dependence of inelasticity is a problem  of great interest
from both theoretical and experimental standpoints \cite{Musulmanbekov}. However, the
experimental data are scarce and, on the theoretical side, the existing models
are largely in conflict with each other even in
explaining a simple aspect as the center-of-mass energy dependence
of the inelasticity \cite{Navarra}, \cite{YH}. For example, the decrease in
inelasticity with energy is advocated by some authors, while others
believe that the inelasticity is an increasing function of energy \cite{Musulmanbekov}.
Hence the problem remains unsolved. Naturally, multiplicity
distributions are connected to inelasticity ones, so
one can study multiplicity distribution features in order to derive
information on the inelasticity behavior. Following this way, we have updated an
existing phenomenological procedure \cite{BeggioMV}, referred as
\emph{Simple One String Model}, which allows simultaneous
description of several experimental data from elastic and inelastic
channels through the Unitarity Equation \cite{BeggioMV}. Thus, based
on the \emph{Simple One String Model} formalism, able to describe
the charged multiplicity distributions from ISR to Collider energies
(30.4 - 900 GeV), we have computed the impact parameter dependence
of inelasticity at fixed center-of-mass energies, $\sqrt{s}$. We have
also inferred information on energy dependence of inelasticity.
The plan of the paper is as follows. In the section 2 we present the
basic formalism of the \emph{Simple One String Model} and the
predictions for $pp/p\overline{p}$ overall multiplicity
distributions compared with the experimental data. In Section 3 we
apply the theoretical framework computing the inelasticity of
hadronic reactions. The final remarks are the content of Section 4.

\section{\emph{Simple One String Model} for Multiplicity Distributions}

The \emph{Simple One String Model} has been discussed in references
\cite{BeggioMV}, \cite{BeggioHama} and in order to define the notation and
also update the model, we shall review the main points here. We work
in impact parameter space, $\emph{b}$, and to guarantee unitarity,
the inelastic cross sections, $\sigma_{in}$, is calculated via the
relation:
\begin{equation}
 \sigma_{in}(s)=\int d^2b\, Gin(s,b)\,,
 \label{sigma}
\end{equation}
where
\begin{equation}
 Gin(s,b)=1-e^{-2\chi_{I}(s,b)}
 \label{sigma}
\end{equation}

is the Inelastic Overlap Function. In this work we have update the
model adopting the complex eikonal function
$\chi_{pp}^{\overline{p}p}{(s,b)}=\chi_{R}{(s,b)}+i\chi_{I}{(s,b)}$,
from Ref. \cite{Block}, as will be discussed in subsection 2.1. In
Ref. \cite{BeggioMV} has been adopted the Henzi Valin parametrization for $Gin(s,b)$. The
probabilities of $n$ particle production, namely multiplicity
distribution, $P_{n}$, is the most general feature of the
multiparticle production processes \cite{Dremin1} and measurements
of charged particle multiplicity distributions have revealed
intrinsic features in $pp$ / $p\overline{p}$ interactions
\cite{Walker}. The multiplicity distribution is defined by the
formula \cite{DreminGary}
\begin{equation}
 P_n(s)={\frac{\sigma_n(s)}{\sum_{n=0}^{\infty}\sigma_{n}(s)}}={\frac{\sigma_n(s)}{\sigma_{in}(s)}},
 \label{PN}
\end{equation}
where $\sigma_n$ is the cross section of an $n$-particle process
(the so-called topological cross section). The charged multiplicity
distribution, in the impact parameter formalism, may be constructed
by summing contributions coming from hadron-hadron collisions taking
place at fixed impact parameter. In this way, the idea of a
normalized multiplicity distribution at each impact parameter $b$ is
introduced \cite{Barshay}. Thus the multiplicity distribution is
written as
\begin{equation}
 P_n(s)={\frac{\sigma_n(s)}{\sigma_{in}(s)}}
       ={\frac{\int d^2b\,G_{in}(s,b)\left[
              {\frac{\sigma_n(s,b)}{\sigma_{in}(s,b)}}
                                     \right]}
              {\int d^2b\,G_{in}(s,b)}}\,,
\end{equation}
where the topological cross section $\sigma_n$ is decomposed into
contributions from each impact parameter $b$ with weight
$G_{in}(s,b)$. In the original formulation \cite{BeggioMV} the
quantity em brackets scales in KNO sense, and the Eq. (4) can be
written as
\begin{equation}
 P_{n}(s)={\frac
          {\int d^2b\,{\frac{G_{in}(s,b)}{<n(s,b)>}}
           \left[<n(s,b)>
           {\frac{\sigma_n(s,b)}{\sigma_{in}(s,b)}}\right]}
          {\int d^2 b\,G_{in}(s,b)}}\,,
 \label{prob}
\end{equation}
where $<n(s,b)>$ is the average number of particles produced at $b$
and $\sqrt{s}$ due to the interactions among hadronic constituents
involved in the collision and, in this model,  $<n(s,b)>$ factorizes
as \cite{BeggioMV}
\begin{equation}
 <n(s,b)> = <N(s)>f(s,b),
 \label{n}
\end{equation}
where $<N(s)>$ is the average multiplicity at $\sqrt{s}$ and
$f(s,b)$ is the so called multiplicity function. Similarly to KNO,
is introduced, for each $b$, the elementary multiplicity
distribution
\begin{equation}
 \psi^{(1)}\left(\frac{n}{<n(s,b)>}\right)
  =<n(s,b)>{\frac{\sigma_n (s,b)}{\sigma_{in}(s,b)}}.
 \label{psi}
\end{equation}
Thus, with Eqs. (\ref{n}) and (\ref{psi}), Eq. (\ref{prob}) becomes
\begin{equation}
 \Phi(s,z)={\frac
  {\int d^2b\,{\frac{G_{in}(s,b)}{f(s,b)}}\,
    \psi^{(1)}\!\left(\frac{z}{f(s,b)}\right)}
  {\int d^2b\,G_{in}(s,b)}},
 \label{Phi}
\end{equation}
where $\Phi(s,z)=<N(s)>P_{n}(s)$ and $z=n/<N(s)>$. Here $z$
represents the usual KNO scaling variable. Now, to obtain the
multiplicity function $f(s,b)$ in terms of the imaginary eikonal
$\chi_{I}$, it has been assumed that
\begin{enumerate}
\item the fractional energy $\sqrt{s^{'}}$ that is
deposited for particle production in a collision at $b$ is
proportional to $\chi_{I}$:
\begin{equation}
\sqrt{s^{'}}=\beta(s)\,\chi_{I}(s,b)\,.
 \label{parten}
\end{equation}
The physical motivation of this equation is that eikonal may be
interpreted as an overlap, on the impact parameter plane, of two
colliding matter distributions \cite{Barshay};

\item  the average
number of produced particles depends on the energy $\sqrt{s^{'}}$ in
the same way as in $e^-e^+$ annihilations, which is approximately
represented by a power law in $\sqrt{s}$ \cite{BeggioMV}
\begin{equation}
 <n(s,b)>=\gamma\,\left(\frac{s^{'}}{s^{'}_{0}}\right)^{A},
 \label{n'}
\end{equation}
\end{enumerate}
where $s^{'}_{0}$=1 GeV$^{2}$. In Ref. \cite{Levin} a power law
energy dependence of multiplicity in both $pp$ and $Pb+Pb$
collisions has been analyzed based on the gluon saturation scenario.
Now, combining Eqs. (\ref{parten}), (\ref{n'}) and (\ref{n}), we have
\begin{equation}
f(s,b)=\frac{\gamma}{<N(s)>}\left[\frac{\beta(s)}{\sqrt{s^{'}_{0}}}\right]^{2A}[\chi_{I}(s,b)]^{2A}=\xi(s)[\chi_{I}(s,b)]^{2A},
\end{equation}

with $\xi(s)$ determined by the usual normalization conditions on
$\Phi$ \cite{BeggioMV} and that serves to determine $\xi(s)$ as an
energy dependent quantity, explicitly \cite{BeggioMV}
\begin{equation}
 \xi(s)=\frac
  {\int d^2 b\,G_{in}(s,b)}
  {\int d^2 b\,G_{in}(s,b)[\chi_{I}(s,b)]^{2A}}.\
 \label{xi}
\end{equation}
Thus, adopting an appropriate parametrization for $G_{in}$ and
$\psi^{(1)}$, as well as an adequate value for $A$, we can test the
formalism embodied in Eq. (\ref{Phi}) and (\ref{xi}), making direct
comparisons with multiplicity distribution data. In the following,
we will discuss the results obtained in the context of our updated
model.

\subsection{Inputs and Results on Multiplicity Distributions}

The \emph{Simple One String Model} is based on the idea of
multiparticle creation due to the interactions between hadronic
constituents in collisions taking place at $b$. It is assumed that in
parton-parton collision there is formation of a string, in which probably
one $q\overline{q}$ has triggered the multitude of the final particles. In previous analysis
the authors \cite{BeggioMV} considered the experimental data
available on $e^-e^+$ annihilations as possible source of
information concerning elementary hadronic interactions and, in this
work, we borrow their results. Specifically, assuming a gamma
distribution normalized to 2,
\begin{equation}
 \psi^{(1)}(z)=2\,\frac{k^k}{\Gamma(k)}z^{k-1}e^{-kz},
 \label{psi1}
\end{equation}
experimental data on $e^-e^+$ multiplicity distributions were
fitted, obtaining \emph{k}=10.775 $\pm$ 0.064
$(\chi^2/N_{DF}=2.61)$. Also, the average multiplicity data in
$e^-e^+$ annihilations were fitted by Eq. (\ref{n'}), giving
$A$=0.258 $\pm$ 0.001 and $\gamma$=2.09 $(\chi^2/N_{DF}=8.89)$ in
the interval 5.1 GeV $\leq \sqrt{s} \leq$ 183 GeV, and $A$=0.198
$\pm$ 0.004 $(\chi^2/N_{DF}=1.7)$ for the set in the interval 10 GeV
$< \sqrt{s} \leq$ 183 GeV, respectively \cite{BeggioMV}. In the
analysis done in Ref. \cite{Levin} the value of $A$=0.11 was
obtained within the gluon saturation picture. The difference between
the values of $A$ is, probably, associated with the different sets
of experimental data used in each analysis. We recall that the value
of $A$=0.11 was obtained by using experimental data for average
multiplicity of hadrons in the gluon and quark jets in $e^-e^+$
annihilation, in the interval of the jet energy between 0.6 $\sim$
32 GeV \cite{Levin}. At the end of the section the \emph{One String
Model} formalism will be tested using the three $A$ values, above
mentioned. Now is needed a parametrization for the eikonal function,
and we have adopted the QCD-inspired complex eikonal from the work
of Block \emph{et al.} \cite{Block} in which the eikonal function is
written as a combination of an even and odd eikonal terms related by
crossing symmetry
$\chi_{pp}^{\overline{p}p}{(s,b)}=\chi^{+}{(s,b)}\pm\chi^{-}{(s,b)}$.
The even eikonal is written as the sum of gluon-gluon, quark-gluon
and quark-quark contributions:
\begin{equation}
\chi^{+}{(s,b)}=\chi_{qq}{(s,b)}+\chi_{qg}{(s,b)}+\chi_{gg}{(s,b)},
 \label{sigma}
\end{equation}
while the odd eikonal, that accounts for the difference between $pp$
and $p\overline{p}$, is parametrized as
\begin{equation}
\chi^{-}{(s,b)}=C^{-}\sum\frac{m_{g}}{\sqrt{s}}e^{i\pi/4}W(b;\mu^{-}).
 \label{sigma1}
\end{equation}
The various parameters and functions involved in last two
expressions are discussed in Ref.\cite{Block}. By fixing the value
of $A$=0.258, $\psi^{(1)}$ given by Eq. (13) with $k$=10.775,
adopting $G_{in}$ from analysis by Block \emph{et al}. and observing
that $\xi(s)$ is obtained by Eq. (12), we have computed the overall
multiplicity distributions arising from $pp/\overline{p}p$
collisions at energies 52.6, 200, 546 and 900 GeV. The theoretical
curves are shown in Figs. 1, 2, 3 and 4 together with the
experimental data. The curves at $pp$-ISR $52.6$ GeV and CERN
$p\overline{p}$ Collider $546$ GeV shows excellent agreement with
the data, Figs. 1 and 3 respectively. At energies $\sqrt{s}$=200 and
900 GeV, also in CERN $p\overline{p}$ Collider, the agreement with
data seems reasonable since the curves agree with experimental
points for $z'\gtrsim 1$, (high multiplicities), Figs. 2 and 4
respectively. In view of the recent results on multiplicity
distributions, in $pp$ collisions at $\sqrt{s}$=0.9, 2.36 and 7 TeV
at the Large Hadron Collider (LHC), reported by CMS Collaboration
\cite{CMS},
\begin{figure}[htt!]
\begin{center}
\includegraphics*[height=9cm,width=8.4cm]{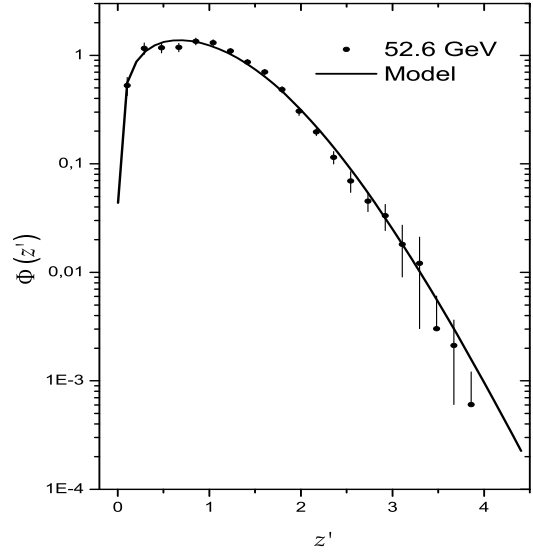}
\caption{Overall scaled multiplicity distribution data for $pp$ at
ISR energy \cite{ABC}, compared to theoretical prediction using the
\emph{Simple One String Model}, Eqs. (8) and (12).} \label{fused445}
\end{center}
\end{figure}

\begin{figure}[ht!]
\begin{center}
\includegraphics*[height=9cm,width=8.4cm]{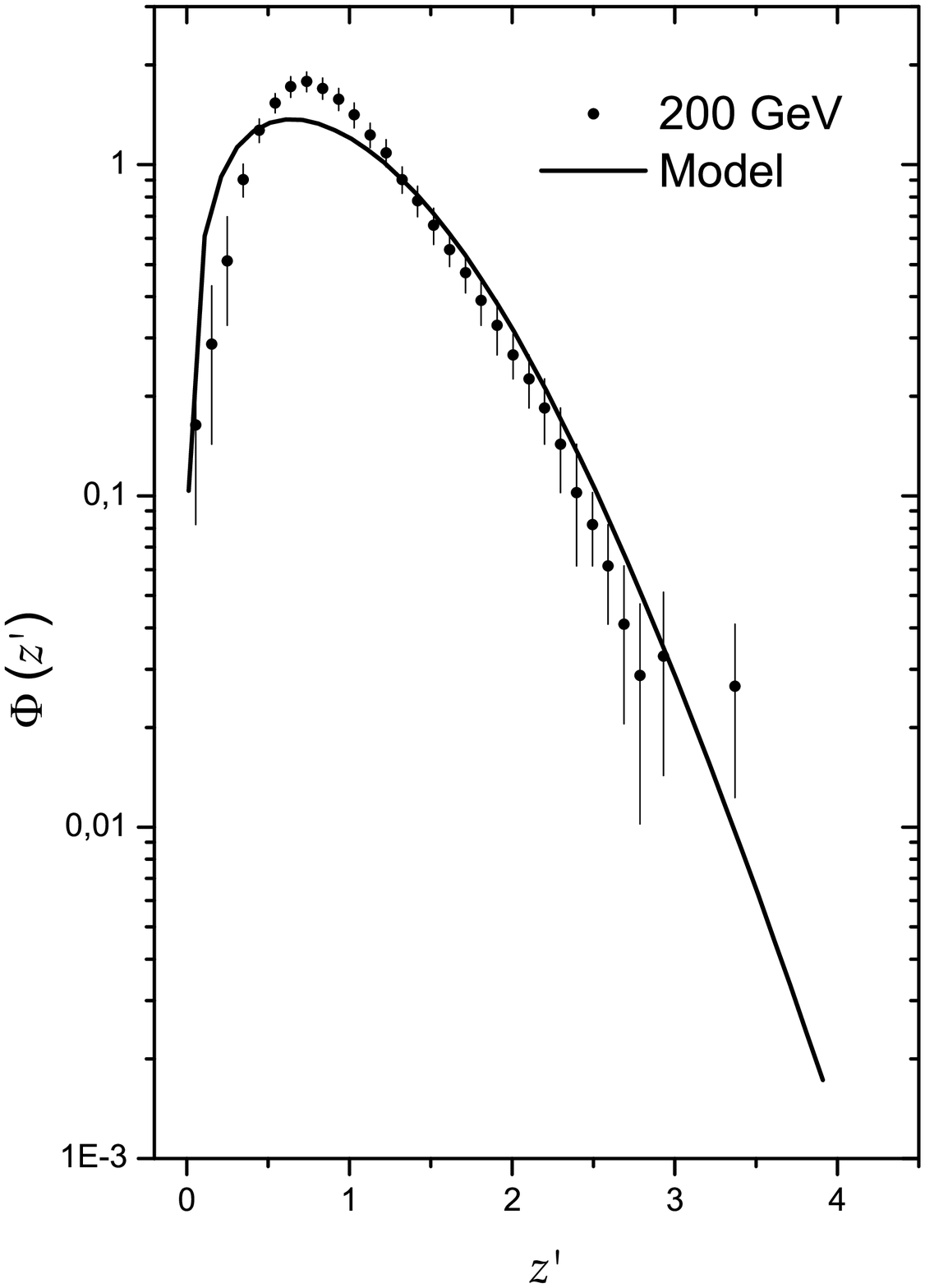}
\caption{Overall scaled multiplicity distribution data for
$p\overline{p}$ at Collider energy \cite{CERN200}, compared to
theoretical prediction using the \emph{Simple One String Model},
Eqs. (8) and (12).} \label{fused445}
\end{center}
\end{figure}
we have also computed the overall multiplicity distributions that
the \emph{One String Model} predicts at LHC energies. The results
are shown in Fig. 5 and we can see violation of KNO scaling, in
qualitative agreement with the result obtained by CMS Collaboration
in pseudorapidity interval of $|\eta|<$2.4, as discussed in Ref.
\cite{CMS}. As mentioned, two fit values of $A$=0.258 and $A$=0.198
have been obtained in the previous study \cite{BeggioMV}, the first
one giving a better account of lower energy data whereas the second
one higher energy data. As before, we have computed the
corresponding hadronic multiplicity distribution by fixing both the
gamma parametrization for $\psi^{(1)}$, Eq. (\ref{psi1}), and the
complex eikonal, Eqs. (\ref{sigma}) and (\ref{sigma1}), and
considered the two parametrizations for the average multiplicity,
$\sim s^{0.258}$ ($\xi(s)$=1.424) and $\sim s^{0.198}$
($\xi(s)$=1.348). The results at 546 GeV are shown in Fig. 6 and,
for $A$=0.198, we can see the disagreement of the theoretical curve
when compared to the data. As pointed out in \cite{BeggioMV} the
parametrization $\sim s^{0.198}$ brings information from data at
high energies, while the parametrization $\sim s^{0.258}$ is in
agreement with data at smaller energies. However, the information
from the $e^-e^+$ average multiplicities at high energies does not
reproduce the overall multiplicity distribution. Hence, by using
$A$=0.256 the output seems to be more consistent with data. In
addition, there is no evidence of gluon saturation at CERN
$p\overline{p}$ Collider 546 GeV, however and as a pedagogical
exercise, we have also computed $\Phi$ considering the value of
$A$=0.11 ($\xi(s)$=1.211), Fig. 6. We would like emphasize that when
the formalism is applied, considering the three values of $A$
(0.258, 0.198 and 0.11) at energies 52.6, 200 and 900 GeV, the
results are essencially the same obtained in Fig. 6. We have
expressed $\Phi$ in terms of modified the scaling variable
$z'=n-N_{o}/<n-N_{o}>$ with $N_{o}$=0.9 representing the average
number of leading particles \cite{BeggioMV}.

\section{Inelasticity}

The concept of inelasticity is essential since it defines the energy
available for particle production in high energy hadronic and
nuclear collisions. However, the impact parameter dependence of the
inelasticity is a problem unsolved. In theoretical works, it is
quite natural to assume that the multiplicity distribution and
inelasticity are connected. Indeed, some authors has defined
multiplicity distributions in terms of inelasticity as
\cite{Musulmanbekov}, \cite{Kadija}
\begin{equation}
 P_n(s)=\int_{0}^{1}P(n|K)P[K(s)]dK,
\end{equation}
where $P[K(s)]$ is the inelasticity distribution and $P(n|K)$ is the
probability of the production of $n$ particles at the given
inelasticity $K$. Thus, based on the connection between multiplicity
and inelasticity, we have explored the parametrization of the
\emph{One String Model} formalism and computed the impact parameter
dependence of inelasticity, as will discuss in next two subsections.

\begin{figure}[ht!]
\begin{center}
\includegraphics*[height=9cm,width=8.4cm]{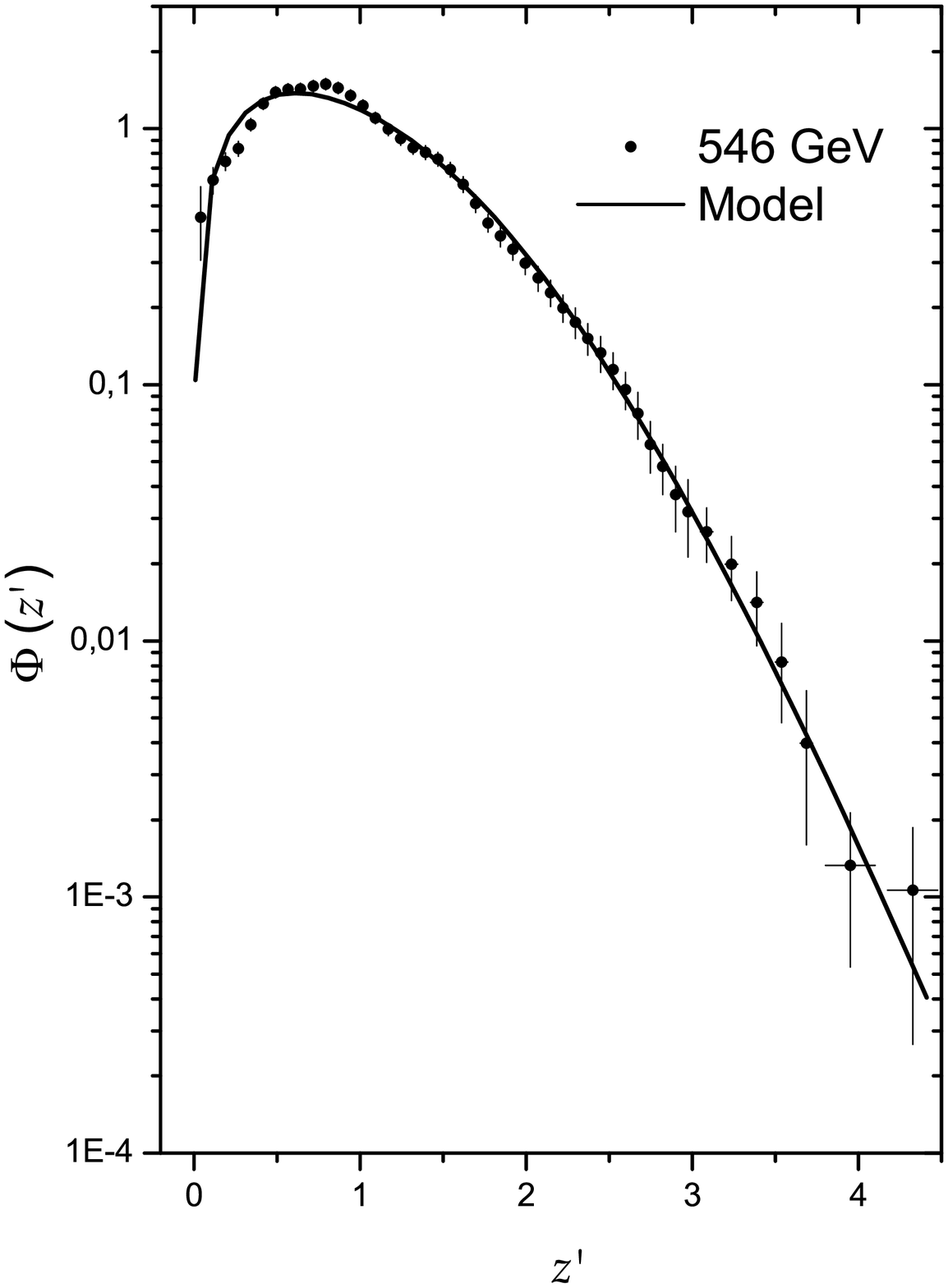}
\caption{Overall scaled multiplicity distribution data for
$p\overline{p}$ at Collider energy \cite{UA51}, compared to
theoretical prediction using the \emph{Simple One String Model},
Eqs. (8) and (12).} \label{fused445}
\end{center}
\end{figure}

\begin{figure}[ht!]
\begin{center}
\includegraphics*[height=9cm,width=8.4cm]{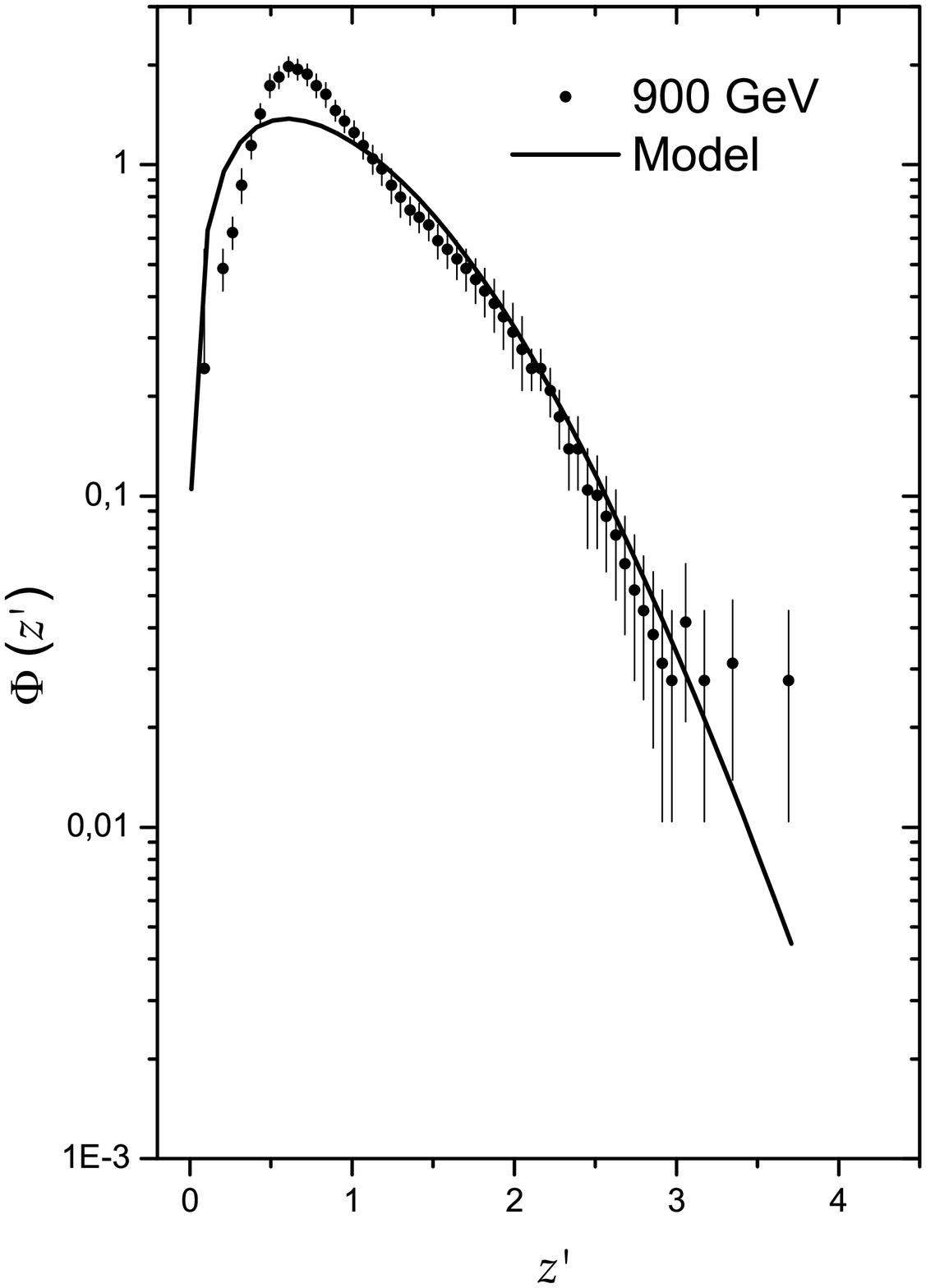}
\caption{Overall scaled multiplicity distribution data for
$p\overline{p}$ at Colider energy \cite{CERN200}, compared to
theoretical prediction using the S\emph{imple One String Model},
Eqs. (8) and (12).} \label{fused445}
\end{center}
\end{figure}

\begin{figure}[ht!]
\begin{center}
\includegraphics*[height=9cm,width=8.4cm]{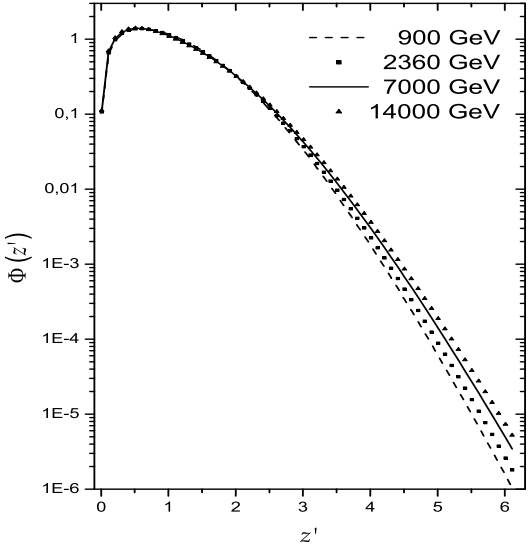}
\caption{Theoretical predictions for overall multiplicity
distribution by using the \emph{Simple One String Model}, Eqs. (8)
and (12), at LHC energies.} \label{fused445}
\end{center}
\end{figure}

\begin{figure}[ht!]
\begin{center}
\includegraphics*[height=9cm,width=8.4cm]{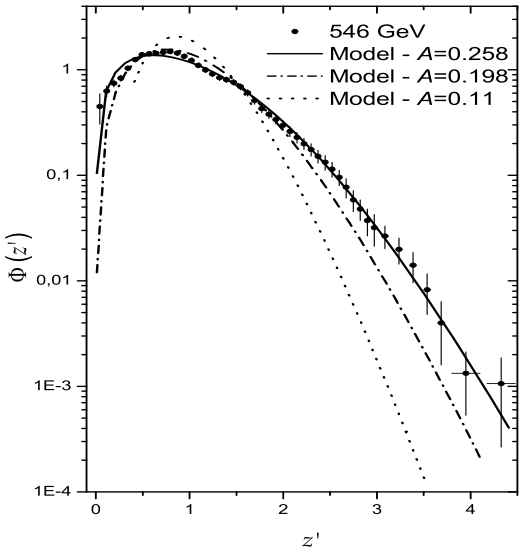}
\caption{Overall scaled multiplicity distribution data for
$p\overline{p}$ at Collider energy \cite{UA51}, compared to
theoretical prediction using the \emph{Simple One String Model},
Eqs. (8) and (12), considering three different values of $A$, Eq.
(\ref{n'}).} \label{fused445}
\end{center}
\end{figure}

\subsection{Similarities between $pp/p\overline{p}$ and
$e^{+}e^{-}$ Collisions}

The idea of a universal hadronization mechanism is not new and
similarities between both processes were indeed observed
\cite{Kadija}, \cite{Basile}, \cite{Fiete}. For example, the average
multiplicities in $pp/p\overline{p}$ and $e^{+}e^{-}$ collisions
become similar when comparisons are made at the same effective
energy for hadron production. In $pp/p\overline{p}$ collisions the
effective energy for particle production, $E_{eff}$, is the energy
left behind by the two leading protons
\begin{equation}
 E_{eff}=(\sqrt{s})_{pp}-(E_{leading,1}+E_{leading,2}),
\end{equation}
or
\begin{equation}
 E_{eff}=(\sqrt{s})_{pp}-2E_{leading},
\end{equation}
in the case of symmetric events. (We recall that $(\sqrt{s})_{pp}$
and $\sqrt{s}$ represents both the center-of-mass energy, however,
in this subsection the notation $(\sqrt{s})_{pp}$ is helpful to
differentiate that from $(\sqrt{s})_{e^{+}e^{-}}$). In $e^{+}e^{-}$
collisions the effective energy for hadron production coincides with
total center-of-mass energy of the beam
\begin{equation}
 E_{eff}=(\sqrt{s})_{e^{+}e^{-}}=2E_{beam}.
\end{equation}
Thus, the same equivalent energy for both $pp/p\overline{p}$ and
$e^{+}e^{-}$ collisions can be written as
\begin{equation}
(\sqrt{s})_{pp}-2E_{leading}=E_{eff}=(\sqrt{s})_{e^{+}e^{-}}.
\end{equation}
For the quantitative estimation of the inelasticity, $K$, we can use
the definition \cite{Musulmanbekov}, \cite{Navarra}
\begin{equation}
 E_{eff}=K(\sqrt{s})_{pp}\Rightarrow K=\frac{E_{eff}}{(\sqrt{s})_{pp}}.
\end{equation}
In the following, we will explore the \emph{One String Model}
formalism, able to describe the multiplicity distributions in wide
interval of energy (30.4 - 900 GeV), to obtain information about inelasticity.

\subsection{Computation of Inelasticity}

The Eq. (9) is a key point of the formalism. Physically, it
corresponds to the energy for hadron production deposited at $b$,
due to the interactions among hadronic constituents involved in the
collision. Thus, and as discussed in last section, the fractional
energy, $\sqrt{s'}$ (Eq. (9)), and the effective energy for hadron
production, $E_{eff}$ (Eq. (21)), represents both the same physical
quantity ($\sqrt{s^{'}}=E_{eff}$). Now, by using the Eq. (9), let us
write the inelasticity, Eq. (21), as a function of $\sqrt{s}$ and
$b$ as
\begin{equation}
2K(s,b)=\frac{\sqrt{s^{'}}}{(\sqrt{s})_{pp}}=\frac{\beta(s)\chi_{I}(s,b)}{\sqrt{s}}.
\end{equation}
The factor 2, in the above equation, is due to the fact that the
multiplicity distributions data are normalized to 2. However, we can
not calculate the $K(s,b)$ until the value of $\beta(s)$ is known.
To estimate $\beta(s)$ we note that the parameter $\xi(s)$, which
is introduced in Eq. (11), is related with $\beta(s)$ by
\begin{equation}
\xi(s)=\frac{\gamma}{<N(s)>}\left[\frac{\beta(s)}{\sqrt{s^{'}_{0}}}\right]^{2A}.
\label{ctebeta}
\end{equation}
We recall that $<N(s)>$ is the average multiplicity at $\sqrt{s}$.
By using the values of $A=0.258$, $\gamma$=$2.09$, discussed in
subsection 2.1, and the values of $<N(s)>$ imputed from experiments
\cite{ABC}, \cite{CERN200} and \cite{UA51}, and also observing that
$\xi(s)$ is obtained from Eq. (12), we have estimated the values of
$\beta$ at various energies. The results are displayed in Table 1.
We can see clearly that $\beta$ increases as the collision energy
also increases. $\beta$ can be parameterized as $\beta(s)=77.48
\sqrt{s} + 0.4168$ with $\chi^{2}/N_{DF}$=1.

\begin{table}[htb!]
\caption{\label{tabpi} $\beta(s)$ estimated values at various
energies. The values of $<N(s)>$ was imputed from Refs. \cite{ABC},
\cite{CERN200} and \cite{UA51}.}
\begin{tabular}{cccc}
\hline
$\sqrt{s}$ GeV & $\xi(s)$ & $<N(s)>$ & $\beta(s)$ GeV\\
\hline
52.6 & $1.612$& $11.55$ & $69.295$\\
200 & $1.517$ & $21.4$ & $203.531$\\
546 & $1.424$ & $27.5$ & $292.729$\\
900 & $1.377$ & $35.6$ & $452.376$\\
2360 & $1.286$ & $-$ & $1061.13$\\
7000 & $1.188$ & $-$ & $2995.08$\\
14000 & $1.130$ & $-$ & $5912.68$\\\hline
\end{tabular}
\end{table}
Now we proceed to compute the impact parameter dependence of
inelasticity and infer some information on its energy dependency.

\subsection{Results and Discussion}

Based on the connection between multiplicity distribution and
inelasticity, we have update and applied the \emph{One String Model}
formalism deriving an expression, Eq. (22), which allows us to study
the impact parameter and energy dependence of inelasticity. Adopting
the Block \emph{et al.} QCD-inspired parametrization for
$\chi_{pp}^{\overline{p}p}{(s,b)}$ \cite{Block} and by using the
estimated values of $\beta(s)$, Table 1, we have applied the Eq.
(22) by fixing the collision energy and computed the inelasticity as
a function of $b$. We show, in Fig. 7, the results from our
analysis. Naturally, the inelasticity decreases as a function of
impact parameter, $b$. The inelasticity behavior is essentially the
same at energies 546 and 900 GeV. At energies 52.6 and 200 GeV, we
can see appreciable difference just in the region of $b \sim 0$
(central collisions). It is interesting to note that, from the range
of collision energy 50 $\sim$ 200 GeV to that one 500 $\sim$ 900
GeV, the inelasticity shows a difference about 60 percent in its
values for $b$ $ \lesssim 0.5$ \emph{fm}. We can also see that, at
fixed $b$, the inelasticity $K$ decreases as $\sqrt{s}$ increases in
the interval 52.6 - 900 GeV. The Eq. (22) depends on the eikonal and
$\beta$ parameter. With the

\begin{figure}[ht!]
\begin{center}
\includegraphics*[height=9cm,width=8.4cm]{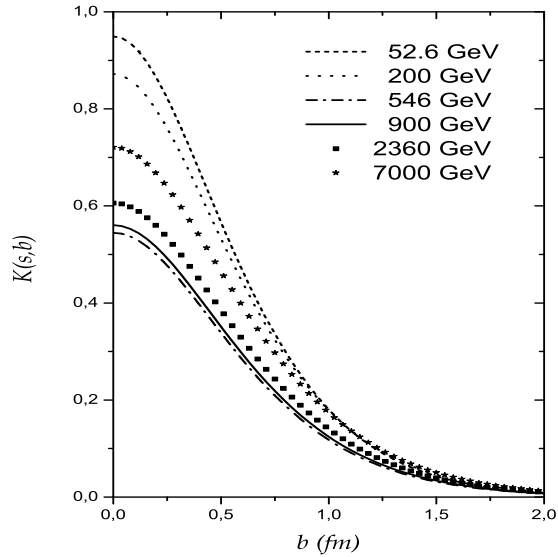}
\caption{Impact parameter dependence of the inelasticity by using
the formula obtained in this work, Eq. (22).} \label{fused445}
\end{center}
\end{figure}

eikonal as determined phenomenologically in \cite{Block} as input,
where high energy cross sections grow with energy as a consequence
of the increasing number of soft partons populating the colliding
particles $(pp/p\overline{p})$, it seems quite natural to expect
that multiparton interactions leads to larger
multiplicities/inelasticities as consequence to the full development
of the gluonic structure. However, looking the same impact parameter
dependence of inelasticity functions at 546 and 900 GeV, Fig. 7, we
would be tempted to conclude that we are observing saturation
effects due gluon recombination in the inelasticity, but there is no
evidence of saturation in this range of energy (52.6 - 900 GeV). We
have also applied our approach at energies $\sqrt{s}$=2.36 and 7 TeV
(LHC), as shown in Fig. 7, and the results suggest that the
inelasticity is an increasing function of energy for the interval
2.36 - 7 TeV. As mentioned before, our main purpose is study
features of multiplicity distributions deriving information on
inelasticity and our analysis is based on the model in which is
assumed that in parton-parton collision there is formation of a
string. Thus, despite some simplifications made in the \emph{One
String Model}, the results seems to be consistent with the
multiplicity distributions data in a wide interval of energy (52.6 -
900 GeV). Hence, the computed inelasticities, in this range of
energy, are reliable results. In counterpart, the results at
$\sqrt{s}$=2.36 and 7 TeV are inconclusive in the context our
analysis, because the \emph{One String Model} probably
underestimates the high multiplicities events due to the lack of the
multicomponent structure in its formulation. In fact, recent results
reported by CMS Collaboration pointed out the importance of a
multicomponent structure in hadron-hadron inelastic interactions, in
agreement with previous experimental results (for details see
\cite{CMS}). Inelasticity also has been studied recently in Ref.
\cite{Wibig}, where, in the context of both Wdowczyk and Wolfendale
model and UHECR data analysis, it was found that the inelasticity
decreases in very high energy interactions, and, in the same work
and by using the modified Feynman scaling formula, the inelasticity
is an increasing function of the energy. It reflects the subtlety of
the theme. We note that at ISR Energies (30-60 GeV), where the
leading particle spectrum could be measured, the inelasticity is
defined to be about 0.5. This value can be identified with $pp$
collision taking place at $b\sim$ 0.6 \emph{fm}, Fig. 7. The
\emph{One String Model} has been used to study the influence on
$\Phi$ considering possible values of $A$ parameter at
$\sqrt{s}$=546 GeV, Fig. 6. In addition, we have also computed
$\Phi$ at energies 52.6, 200 and 900 GeV using the different values
of $A$ (0.258, 0.198 and 0.11) and the results, in each energy, are
similar with that obtained in Fig. 6. Finally, we emphasize that the
curves in Figs. 1-4 has not been fit to data, except for the values
of $A$ and $k$ (fixed) no experimental information about
multiplicity distribution has gone into the calculation. Hence, the
energy evolution of the multiplicity distributions, from ISR to
Collider ($30 \sim 900\,$GeV), is correctly reproduced by changing
only the Overlap Function, Eq.(2), without changing the underlying
elementary interaction, in agreement with what could be expected
from QCD.

\section{Concluding Remarks}

Being the impact parameter $b$ an essential variable in a
geometrical description of hadronic collisions, we have investigated
the $b$ dependence of inelasticity and also inferred some
information on its energetic behavior. By using a geometrical model
we have derived an expression for $K$ based on the hypothesis of
connection between multiplicity distribution and inelasticity. We
have adopted the Block \emph{et al.} model in our analysis, where
the eikonal functions $\chi_{qq}{(s,b)}$ and $\chi_{qg}{(s,b)}$ are
needed to describe the lower energy forward data, while
$\chi_{gg}{(s,b)}$ contribution dominates at high energy and
determines the asymptotic behavior of cross sections. We believe
that the same impact parameter dependence of the inelasticity at
energies 546 and 900 GeV can have important implications for the
underlying gluon-gluon dynamics. In fact, we are testing our
formalism by using the QCD Eikonal Model in which the gluon may
develop a dynamical mass \cite{Luna}, \cite{Luna1}. At energies
$\sqrt{s}$=2.36 and 7 TeV the \emph{One String Model}
parametrization can not be tested, hence the results do not allow
for any conclusion. Finally, the results suggest that there are
relationships between the inelasticity and the eikonal function.

\begin{ack}
I am grateful to E.G.S. Luna for several instructive discussions.
\end{ack}


\end{document}